\documentclass[aps]{revtex4}
\usepackage{amssymb,epsf}
\usepackage{amsfonts}
\usepackage{amsmath}

\begin{document}

\title{Energy dependent topological adS black holes in Gauss--Bonnet
Born--Infeld gravity}
\author{S. H. Hendi$^{1,2}$\footnote{email address: hendi@shirazu.ac.ir}, H. Behnamifard$^{1}$ and B.
Bahrami-Asl$^{1}$}
\affiliation{$^1$Physics Department and Biruni
Observatory, College of Sciences, Shiraz
University, Shiraz 71454, Iran \\
$^2$Research Institute for Astrophysics and Astronomy of Maragha (RIAAM),
P.O. Box 55134-441, Maragha, Iran }

\begin{abstract}
Employing higher curvature corrections to Einstein--Maxwell
gravity has garnered a great deal of attention motivated by the
high energy regime in quantum nature of black hole physics. In
addition, one may employ gravity's rainbow to encode quantum
gravity effects into the black hole solutions. In this paper, we
regard an energy dependent static spacetime with various
topologies and study its black hole solutions in the context of
Gauss--Bonnet Born--Infeld (GB--BI) gravity. We study
thermodynamic properties and examine the first law of
thermodynamics. Using suitable local transformation, we endow the
Ricci--flat black hole solutions with a global rotation and study
the effects of rotation on thermodynamic quantities. We also
investigate thermal stability in canonical ensemble through
calculating the heat capacity. We obtain the effects of various
parameters on the horizon radius of stable black holes. Finally,
we discuss second order phase transition in the extended phase
space thermodynamics and investigate the critical behavior.
\end{abstract}

\maketitle

\section{Introduction}

James Clark Maxwell is one of the physicists who played the most
decisive part in developing the electrodynamics. He collected and
modified four equations that are known as the Maxwell's equations,
which are the foundation of classical electromagnetism. Maxwell
electrodynamics contains various problems which motivate one to
consider nonlinear electrodynamics (NED). One of the main
problematic aspects of Maxwell theory is having an infinite
self--energy for a point--like charge. To remove such divergency,
Born and Infeld introduced one of the interesting theories of NED
which is identified as Born--Infeld (BI) theory \cite{BI1,BI2}.
Recently, such nonlinear theory has acquired a new impetus, since
it can be naturally
arisen in the low--energy limit of the open string theory \cite%
{BIstring1,BIstring2,BIstring3,BIstring4,BIstring5}.

On the gravitational point of view, Einstein gravity is a
traditional theory of general relativity. Although such theory is
successful for various aspects and is consistent with some
observational evidences, it has some problems in higher curvature
regimes. The natural extension of Einstein gravity is to consider
the higher curvature gravity which is appeared as next-to-leading
term in heterotic string effective action. Keeping quadratic
curvature terms (and ignoring higher curvature terms) in such
stringy effective action \cite
{GBstring1,GBstring2,GBstring3,GBstring4,GBstring5,GBstring6,GBstring7}
leads to the Gauss--Bonnet (GB) Lagrangian which is a topological
invariant in four dimensions. Various aspects of GB gravity and
its thermodynamic properties have been investigated in several
papers \cite
{GBpapers1,GBpapers2,GBpapers3,GBpapers4,GBpapers5,GBpapers6,GBpapers7,
GBpapers8,GBpapers9,GBpapers10,GBpapers11,GBpapers12,GBpapers13,GBpapers14,GBpapers15,GBpapers16,GBpapers17,GBpapers18,GBpapers19}.

The electromagnetic and gravitational fields near the charged
black holes are very strong, and hence, both the nonlinear
electromagnetic effects and higher curvature gravitational terms
should be taken into consideration. In other words, Maxwell theory
and Einstein gravity can be, respectively, considered as
approximations of NED theory and higher curvature gravity in the
weak fields limit. Regarding the mentioned subjects, one is
motivated to provide an analytically solvable theory which
contains higher curvature corrections of gauge--gravity theories.
Among generalizations of the Einstein--Maxwell action to a higher
curvature gauge--gravity theory, the GB--BI gravity has some
particular interests because it is a ghost-free theory in the
gravitational side and a divergence-free field in the
electromagnetic side. In addition, both GB and BI theories emerge
in the effective low-energy
action of string theory \cite%
{BIstring1,BIstring2,BIstring3,BIstring4,BIstring5,GBstring1,GBstring2,
GBstring3,GBstring4,GBstring5,GBstring6,GBstring7}. The influences of GB
gravity and BI theory have been investigated in various physical phenomena
\cite{GBeffects1A,GBeffects1B,GBeffects1C,GBeffects2A,GBeffects2B,
GBeffects3,GBeffects4,GBeffects5A,GBeffects5B,GBeffects6A,GBeffects6B,
BIsuperconductor1,BIsuperconductor2,BIsuperconductor3,BIHydrogen1,
BIHydrogen2,BIHydrogen3,BIHawRad,BIHydrodynamics,BICosmology1,BICosmology2,BICosmology3}%
. Moreover, in the subject of black hole physics, various properties of
GB--BI solutions were studied (for a very incomplete list of references, see
\cite{GBBI1A,GBBI1B,GBBI2}).

On the other hand, regarding high energy regime of gravitational
physics, it is important to find the UV generalization of general
relativity. There are various attempts to obtain the UV completion
of general relativity, in which they should reduce to the general
relativity in the IR limit. One of approaches for obtaining the UV
completion of general relativity is based on the deformation of
the usual energy-momentum dispersion relation in the UV limit,
which is called gravity's rainbow \cite{Grb}. Considering
gravity's rainbow, one finds that the geometry of spacetime is
made energy dependent
through the introduction of rainbow functions \cite%
{Hendi-FaizalA,Hendi-FaizalB}.

On the other hand, one of the fundamental challenging subjects in
cosmology is big bang singularity. It is shown that nonlinear
electrodynamics can remove both the black hole and big bang
singularities
\cite%
{Nonsingular-nonlinearA,Nonsingular-nonlinearB,Nonsingular-nonlinearC,Nonsingular-nonlinearD,Nonsingular-nonlinearE}%
. Besides, one can construct a nonsingular universe by considering an
effective energy dependent spacetime \cite%
{NonsingularRainbow1,NonsingularRainbow2,NonsingularRainbow3,NonsingularRainbow4,NonsingularRainbow5}%
. In addition, an isotropic perfect fluid model is considered in the context
of gravity's rainbow to obtain a nonsingular bouncing universe \cite%
{NonsingularCosmology1,NonsingularCosmology2}. Moreover, the
absence of singular black holes at the LHC is explained with the
formalism of gravity's rainbow \cite{NonsingularLHC}. In other
words, gravity's rainbow is an effective approach for helping us
to avoid the mentioned singularities.

It was shown that for a certain choice of rainbow functions, there is a
close relation between gravity's rainbow and Horava-Lifshitz gravity \cite%
{Rb-Lf}. In addition, one may regard string theory motivation of considering
gravity's rainbow. The background fluxes in string theory produce a
noncommutative deformation of the geometry \cite{Noncom1A,Noncom1B}, and
such noncommutativity has also been used to motivate one of the most
important classes of rainbow functions in gravity's rainbow \cite%
{Noncom2A,Noncom2B}. Also, applying the spontaneous breaking of the Lorentz
symmetry, one should use a deformation of the usual energy-momentum
relations, and such modification can be used as a motivation for introducing
gravity's rainbow \cite{SSLB}.

It is worthwhile to mention that if the energy of the particle,
$E$, was just a non-dynamical parameter, one could gauge it away
by rescaling. However, it dynamically depends on the coordinates
and it does break original the diffeomorphisms symmetry of the
full metric. In fact, even the local symmetry in gravity's rainbow
is not Lorentz symmetry, as it is based on modified dispersion
relation. An explicit form of coordinate dependence of the energy
for a particular solution was discussed in Ref. \cite{Rb-Lf}.
Although it is hard to find explicit dependence of the energy on
the coordinates for various solutions, it is important to know
that $E$ is a dynamical parameter and cannot be gauged away by
rescaling.

After discovering the fact that the Hawking temperature and
entropy are, respectively, proportional to black hole's surface
gravity and the area of its event horizon, investigation of black
hole thermodynamics becomes a very interesting subject. In
addition, regarding the cosmological constant as a dynamical
pressure, one can study the critical behavior of a system in the
extended phase space. In this regard, one finds an analogy between
van der Waals liquid--gas system and the black hole as a
thermodynamical system
(for an incomplete references, see \cite%
{PressureLambda1,PressureLambda2,PressureLambda3,PressureLambda4,
PressureLambda5,PressureLambda6,PressureLambda7,PressureLambda8,
PressureLambda9,PressureLambda10,PressureLambda11}).

The main goal of this paper is to find topological black hole
solutions of the GB--BI gravity with an energy dependent
spacetime. Although the system seems to be complex, as we
mentioned, considering each item is based on a fundamental theory.
It is worth mentioning that all higher curvature gravity, higher
dimensional physics, higher curvature terms in gauge theory, and
the existence of upper limit for the energy of a particle can be
originated from heterotic string theory and modified dispersion
relation, which are two fundamental acceptable theories. In this
paper, we consider most important theory as well as simplest ones.
In other words, between all higher curvature gravity (such as
higher orders of Lovelock gravity and also $F(R)$ gravity
theories), we select Gauss-Bonnet gravity which is the simplest
second order derivative theory. On the other hand, regarding
various models of nonlinear electrodynamics, we take into account
the most interesting theory which has acceptable results.
Regarding this theory, we could solve some problematic subjects
from field theory to cosmological scales. Furthermore, regarding
an energy dependent line element is the simplest way for
considering the energy of particles on the geometry of spacetime.
Hence, in this paper, we try to combine some interesting subjects,
while we take care for the simplicity. We also discuss
thermodynamic properties of such static topological black holes
and generalize Ricci--flat adS solutions to rotating black branes.
We show that how thermodynamical properties and possible phase
transition of black hole can be affected by the higher-order
curvature terms of gauge/gravity fields and the energy at which
spacetime is probed. In other words, we study thermal stability
and analyze the effects of GB, BI, rotation parameters and rainbow
functions on stability criteria.

\section{Topological Black Hole Solutions \label{Fiel}}

In this section, we are going to obtain $d-$dimensional black hole
solutions with various horizon topologies. In the context of
gravity's rainbow, one may introduce a one-parameter family of
energy dependent orthonormal frame field. As a result,
one-parameter family of energy dependent metric has been created
such as $g^{\mu \nu }(\varepsilon )=e_{a}^{\mu }(\varepsilon
)e^{a\nu }(\varepsilon )$ where $\varepsilon =E/E_{p}$ is the
energy ratio \cite{Grb}. In this relation, $E$ and $E_{p}$, are
respectively, the energy of test particle and Planck energy. At
the first step, we consider the following energy dependent
spherically symmetric line element:
\begin{equation}
d\tau ^{2}=-ds^{2}=A(r,\varepsilon)dt^{2}-
B(r,\varepsilon)dr^{2}-C(r,\varepsilon) d\Omega _{k}^{2} ,  \label{Metric}
\end{equation}
where $A(r,\varepsilon)$, $B(r,\varepsilon)$ and $C(r,\varepsilon)$ are
functions of energy and radial coordinate. In addition, $d\Omega _{k}^{2}$
indicates the Euclidian metric of a $(d-2)-$dimensional hypersurface with
constant curvature $(d-2)(d-3)k$ and volume $V_{d-2}$, where $k$ is a
constant. Hereafter, we take $k=\pm 1,0$, without loss of generality. For
more clarifications, we present the explicit form of $d\Omega _{k}^{2}$ as
\begin{equation}
d\Omega _{k}^{2}=\left\{
\begin{array}{cc}
d\theta _{1}^{2}+\sum\limits_{i=2}^{d-2}\prod\limits_{j=1}^{i-1}\sin
^{2}\theta _{j}d\theta _{i}^{2} & k=1 \\
\sum\limits_{i=1}^{d-2}d\phi _{i}^{2} & k=0 \\
d\theta _{1}^{2}+\sinh ^{2}\theta _{1}(d\theta
_{2}^{2}+\sum\limits_{i=3}^{d-2}\prod\limits_{j=2}^{i-1}\sin ^{2}\theta
_{j}d\theta _{i}^{2}) & k=-1%
\end{array}%
\right. .  \label{dOmega}
\end{equation}

The energy dependent metric functions are considered to make a connection
between these new tetrad fields and the usual frame fields $\tilde{e}%
_{a}^{\mu }$ in general relativity, $f(\varepsilon )e_{0}^{\mu }(\varepsilon
)=\tilde{e}_{0}^{\mu }$ and $g(\varepsilon )e_{i}^{\mu }(\varepsilon )=%
\tilde{e}_{i}^{\mu }$, where $i$ is the spatial index and $f(\varepsilon )$
and $g(\varepsilon )$ are rainbow functions. We also expect to recover the
general relativity in the IR limit, as $\lim_{\varepsilon \rightarrow
0}f(\varepsilon )=\lim_{\varepsilon \rightarrow 0}g(\varepsilon )=1$.
Hereafter, we use the following separation of variable for simplicity and
also consistency with the energy dependent modified dispersion relation
\begin{equation}
A(r,\varepsilon)=\frac{\psi (r)}{f(\varepsilon )^{2}},\;\;\;\;B(r,%
\varepsilon)=\frac{1}{\psi(r)g(\varepsilon)^2},\;\;\;\;C(r,\varepsilon)=%
\frac{r^2}{g(\varepsilon)^2},  \label{change}
\end{equation}
where in the IR limit the Schwarzschild like metric is recovered.
Although there are different models of the rainbow functions with
various phenomenological motivations (see Ref.
\cite{Hendi-FaizalA} for more details), we try to present all
analytical relations with closed form of rainbow functions. This
general form may help us to follow the trace of temporal
($f(\varepsilon )$) and spatial ($g(\varepsilon )$) rainbow
functions, separately. But for plotting the diagrams and other
numerical analysis, we have to choose a class of rainbow functions
with adjusting free parameters.

In order to discuss geometric properties, we should calculate scalar
curvatures. Regarding the metric (\ref{Metric}) with Eq. (\ref{change}), it
is easy to show that the Ricci and Kretschmann scalars are, respectively,
\begin{eqnarray}
R &=&g(\varepsilon )^{2}\left[ -\psi ^{\prime \prime }(r)-2(d-2)\frac{\psi
^{\prime }(r)}{r}+(d-2)(d-3)\frac{k-\psi (r)}{r^{2}}\right] ,  \label{R} \\
R_{\mu \nu \rho \sigma }R^{\mu \nu \rho \sigma } &=&g(\varepsilon )^{4}\left[
\psi ^{\prime \prime 2}(r)+2(d-2)\left( \frac{\psi ^{\prime }(r)}{r}\right)
^{2}+2(d-2)(d-3)\left( \frac{k-\psi (r)}{r^{2}}\right) ^{2}\right] ,
\label{R2}
\end{eqnarray}
where prime and double prime are the first and second derivatives
with respect to $r$ , respectively. Calculating the other
curvature invariants (such as Ricci square, Weyl square and so
on), we find that they are functions of $\psi ^{\prime \prime }$,
$\psi ^{\prime }/r$ and $(k-\psi )/r^{2}$ and therefore, it is
sufficient to study the Kretschmann scalar for investigating the
spacetime curvature. Regarding Eqs. (\ref{R}) and (\ref{R2}), one
finds the considerable effect of rainbow function on all terms of
curvature scalars. This indicates that the curvature of spacetime
can be related to the energy at which spacetime is probed.

Since gravity's rainbow has considerable effects at UV regime, it
may be motivated to consider high energy and quantum corrections
of gravity and gauge fields. In this regard, we are interested in
the GB gravity coupled to a nonlinear $U(1)$ gauge field.

In addition, we should note that the constants may depend on the
scale at which a theory is probed \cite{renom}. Such dependency is
based on the renormalization group flow. Also, according to the
supergravity solutions, one may expect to consider energy
dependent constants. As a result, it is interesting to take into
account energy dependent constants. The action under study is
\begin{equation}
I_{G}=-\frac{1}{16\pi }\int_{\mathcal{M}}d^{d}x\sqrt{-g}\left[
R-2\Lambda(\varepsilon) +\alpha(\varepsilon) L_{GB}+L(\mathcal{F})\right]
\label{Ig}
\end{equation}%
where $\Lambda(\varepsilon) $ and $\alpha(\varepsilon)$ are, respectively,
the energy dependent cosmological constant and GB parameter. In addition, $%
L_{GB}$ and $L(\mathcal{F})$ refer to the Lagrangians of GB and BI
theories, which can be written as
\begin{eqnarray}
&&L_{GB}=R_{\mu \nu \gamma \delta }R^{\mu \nu \gamma \delta }-4R_{\mu \nu
}R^{\mu \nu }+R^{2},  \label{LGB} \\
&&L(\mathcal{F})=4\beta(\varepsilon) ^{2}\left( 1-\sqrt{1+\frac{\mathcal{F}}{%
2\beta(\varepsilon) ^{2}}}\right) ,  \label{L(F)}
\end{eqnarray}%
where $\beta(\varepsilon) $ is called the BI parameter, $\mathcal{F}=F_{\mu
\nu }F^{\mu \nu }$ is the Maxwell invariant, in which $F_{\mu \nu }=\partial
_{\lbrack \mu }A_{\nu ]}$ is the Faraday tensor and $A_{\mu }$ is the gauge
potential. Taking into account the action (\ref{Ig}) and using the
variational principle, we can obtain the following field equations
\begin{eqnarray}
G_{\mu \nu }^{(0)}+G_{\mu \nu }^{(1)}+ &&G_{\mu \nu }^{(2)}=\frac{1}{2}L(%
\mathcal{F})g_{\mu \nu }-2L_{\mathcal{F}}F_{\mu \lambda }F_{\nu }^{\;\lambda
},  \label{Geq} \\
&&\nabla _{\mu }\left( L_{\mathcal{F}}F^{\mu \nu }\right) =0,  \label{Maxeq}
\end{eqnarray}%
where $L_{\mathcal{F}}=\frac{dL(\mathcal{F})}{d\mathcal{F}}$, $G_{\mu \nu
}^{(0)}=\Lambda(\varepsilon) g_{\mu \nu }$, $G_{\mu \nu }^{(1)}$ is the
Einstein tensor and $G_{\mu \nu }^{(2)}$\ is the divergence-free symmetric
tensor of GB contribution
\begin{equation}
G_{\mu \nu }^{(2)}=-\alpha(\varepsilon) \left( 4R^{\rho \sigma }R_{\mu \rho
\nu \sigma }-2R_{\mu }^{\ \rho \sigma \lambda }R_{\nu \rho \sigma \lambda
}-2RR_{\mu \nu }+4R_{\mu \lambda }R_{\text{ \ }\nu }^{\lambda }+\frac{L_{GB}%
}{2}g_{\mu \nu }\right) .  \label{Heq1}
\end{equation}

Regarding the metric (\ref{Metric}) with the field equations (\ref{Geq}) and
(\ref{Maxeq}), we can obtain energy dependent metric function. At first, we
start with a consistent gauge potential $A_{\mu }$ with the following form
\begin{equation}
A_{\mu }=\frac{h(r)}{f(\varepsilon )}\delta _{\mu }^{0}  \label{Amu1}
\end{equation}

The functional form of $h(r)$ can be obtained from Eq. (\ref{Maxeq})
\begin{equation}
h(r)=-\frac{q(\varepsilon)}{(d-3)r^{d-3}}\ \mathcal{H},  \label{h(r)}
\end{equation}%
where
\begin{eqnarray}
\mathcal{H} &=&{}_{2}F_{1}\left( \left[ \frac{1}{2},\frac{d-3}{2d-4}\right] ,%
\left[ \frac{3d-7}{2d-4}\right] ,-\eta \right) ,  \label{hyper} \\
\eta &=&\frac{g(\varepsilon )^{2}q(\varepsilon)^{2}}{\beta(\varepsilon)
^{2}r^{2d-4}},  \label{eta}
\end{eqnarray}%
in which $q(\varepsilon)$ is an integration constant proportional
to the electric charge and ${}_{2}F_{1}\left( \left[ a,b\right] ,\left[ c%
\right] ,x\right) $ is hypergeometric function. Using series expansion for
large $r$ (or large $\beta(\varepsilon) $), we find that the gauge potential
is well-behaved, asymptotically
\begin{equation}
\left. h(r)\right\vert _{\text{Large }r}=-\frac{q(\varepsilon)}{(d-3)r^{d-3}}%
+\frac{g(\varepsilon )^{2}q(\varepsilon)^{3}}{2(3d-7)\beta(\varepsilon)
^{2}r^{3d-7}}+O\left( \frac{1}{\beta(\varepsilon) ^{4}r^{5d-11}}\right).
\label{asymph}
\end{equation}

Eq. (\ref{asymph}) confirms that for large values of $r$ (or $%
\beta(\varepsilon) $), the dominant (first) term of $h(r)$ is the same as
the gauge potential in $d$-dimensional linear Maxwell theory and other terms
are the Maxwell corrections. It is also notable that the rainbow function
affects the correction terms.

Inserting metric (\ref{Metric}) into the gravitational field equations with
obtained gauge potential, one finds the following nonzero components of Eq. (%
\ref{Geq})
\begin{eqnarray}
e_{1} &=&\left[ 2\alpha(\varepsilon) (d-4)\frac{g(\varepsilon )^{2}}{r^{2}}%
\left[ \psi (r)-k\right] -\frac{1}{d-3}\right] \frac{\psi ^{\prime }(r)}{r}-%
\frac{\left[ \psi (r)-k\right] }{r^{2}}  \notag \\
&&+(d-4)(d-5)\alpha(\varepsilon) g(\varepsilon )^{2}\left[ \frac{\psi (r)-k}{%
r^{2}}\right] ^{2}+\frac{4\beta(\varepsilon) ^{2}(1-\sqrt{1+\eta }%
)-2\Lambda(\varepsilon) }{(d-2)(d-3)g(\varepsilon )^{2}}=0,  \label{e1} \\
&&  \notag \\
&&  \notag \\
e_{2} &=&\left[ \frac{1}{(d-3)(d-4)}-2\alpha(\varepsilon) g(\varepsilon )^{2}%
\frac{\left[ \psi (r)-k\right] }{r^{2}}\right] \psi ^{\prime \prime
}(r)-2\alpha(\varepsilon) g(\varepsilon )^{2}\left( \frac{\psi ^{\prime }(r)%
}{r}\right) ^{2}  \notag \\
&&-2\alpha(\varepsilon) g(\varepsilon )^{2}\left[ 2(d-5)\frac{\left[ \psi
(r)-k\right] }{r^{2}}-\frac{1}{\alpha(\varepsilon) (d-4)g(\varepsilon )^{2}}%
\right] \frac{\psi ^{\prime }(r)}{r}+\frac{\left[ \psi (r)-k\right] }{r^{2}}
\notag \\
&&-(d-5)(d-6)\alpha(\varepsilon) g(\varepsilon )^{2}\frac{\left[ \psi (r)-k%
\right] ^{2}}{r^{d-3}}+\frac{2\Lambda(\varepsilon) -4\beta(\varepsilon)
^{2}\left( 1-\frac{1}{\sqrt{1+\eta }}\right) }{(d-3)(d-4)g(\varepsilon )^{2}}%
=0.  \label{e2}
\end{eqnarray}

It is a matter of calculation to show that the following metric
function satisfies all field equations, simultaneously
\begin{equation}
\psi (r)=k+\frac{r^{2}}{2\alpha(\varepsilon) (d-4)(d-3)g(\varepsilon )^{2}}%
\left( 1-\sqrt{\digamma (r)}\right)  \label{fr}
\end{equation}%
with%
\begin{eqnarray}
\digamma (r) &=&1+\frac{8(d-3)(d-4)\alpha(\varepsilon) }{(d-1)(d-2)}\left(
\Lambda(\varepsilon) +\frac{(d-1)(d-2)m(\varepsilon)}{2r^{d-1}}-\Theta
(r)\right) ,  \label{Xi} \\
\Theta (r) &=&2\beta(\varepsilon) ^{2}\left[ \left( 1-\sqrt{1+\eta }\right) +%
\frac{d-2}{d-3}\eta \mathcal{H}\right] ,  \label{Theta}
\end{eqnarray}%
where $m(\varepsilon)$ is an integration constant. For large values of
nonlinearity parameter ($\beta(\varepsilon) \rightarrow \infty $), this
solution reduces to GB-Maxwell gravity's rainbow black hole \cite%
{Hendi-FaizalA}. In addition, obtained solutions reduce to GB-BI black holes
for IR limit $f(\varepsilon )=g(\varepsilon )=1$.

Using series expansion of metric function for large distance, we
find that the dominant term is proportional to cosmological
constant. Therefore, obtained solutions are asymptotically (a)dS
with an effective cosmological constant (see Ref. \cite{GBBI2} for
more details). Depending on the free parameters, one finds that
the function $\digamma (r)$ may be positive, zero or negative. In
order to obtain real (physical) solutions, we can use one of the
following two methods. First, we can limit ourselves to the set of
parameters, which lead to non-negative $\digamma (r)$ for $0\leq
r<\infty $. Second, we can define $r_{R}$ as the largest root of
$\digamma (r=r_{R})=0 $, in which $\digamma (r)$ is positive for
$r>r_{R}$. One can use suitable coordinate transformation
($r\longrightarrow r^{\prime }$) to obtain real
solutions for $0\leq r^{\prime }<\infty $ (see the last reference in \cite%
{GBBI2} for more details). Hereafter, we follow the first method.

\begin{figure}[tbp]
\epsfxsize=9cm \epsffile{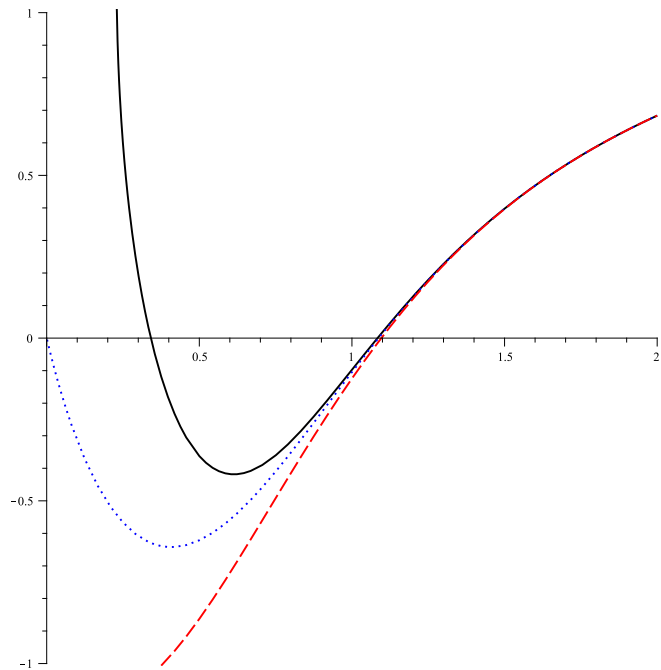}
\caption{$\protect\psi(r)$ versus $r$ for $k=1$, $d=5$, $g(\protect%
\varepsilon)=f(\protect\varepsilon)=0.95$, $\Lambda(\protect%
\varepsilon)=-0.1$, $\protect\alpha(\protect\varepsilon)=0.1$, $q(\protect%
\varepsilon)=1$, and $\protect\beta(\protect\varepsilon)>\protect\beta_{c}(%
\protect\varepsilon)$ (continuous line: timelike singularity), $\protect\beta%
(\protect\varepsilon)=\protect\beta_{c}(\protect\varepsilon) \approx 1.032$
(dotted line: null singularity) and $\protect\beta(\protect\varepsilon)<%
\protect\beta_{c}(\protect\varepsilon)$ (dashed line: spacelike
singularity). }
\label{SingularityType}
\end{figure}

Now, we focus on obtaining black hole solutions. Inserting Eq.
(\ref{fr}) in Eqs. (\ref{R}) and (\ref{R2}), we find that both the
Kretschmann and Ricci scalars diverge at $r=0$ and are finite for
$r\neq 0$. Since there is at least one real positive root for the
metric function, the mentioned singularity can be covered with an
event horizon, and therefore, the solutions can be interpreted as
black holes. In order to discuss the type of horizon, we can
follow Refs. \cite{HendiNON1,HendiNON2}. It was shown that there
is a critical value for nonlinearity parameter, $\beta
_{c}(\varepsilon)$, in which for $\beta(\varepsilon) <\beta
_{c}(\varepsilon) $, the horizon geometry of nonlinear charged
solutions behaves like
Schwarzschild solutions and singularity is spacelike. For $%
\beta(\varepsilon) >\beta _{c}(\varepsilon)$, the horizon geometry is the
same as Reissner--Nordstr\"{o}m black hole and the singularity is timelike.
The mentioned singularity is null for $\beta(\varepsilon) =\beta
_{c}(\varepsilon)$ (see Fig. \ref{SingularityType} for more details).

\section{Thermodynamics: \label{Thermodynamics}}

In this section, we discuss thermodynamical behavior of the black hole
solutions. At first we calculate the Hawking temperature. To do so, we can
use the surface gravity interpretation or follow the absence of conical
singularity at the horizon in the Euclidean sector of the black hole
solutions. Both methods lead to the following temperature%
\begin{equation}
T=\frac{g(\varepsilon )}{4\pi f(\varepsilon )}\left. \frac{d\psi (r)}{dr}%
\right\vert _{r=r_{+}}.  \label{T}
\end{equation}%
After some manipulations, we obtain
\begin{eqnarray}
T &=&\frac{r_{+}\left[ \frac{4\beta(\varepsilon) ^{2}\left[ 1-\sqrt{1+\eta
_{+}}\right] -2\Lambda(\varepsilon) }{(d-2)}+\frac{(d-3)kg(\varepsilon
)^{2}\left( 1+\frac{k\alpha(\varepsilon) g(\varepsilon )^{2}}{r_{+}^{2}}%
(d-5)(d-4)\right) }{r_{+}^{2}}\right] }{4\pi f(\varepsilon )g(\varepsilon )%
\left[ 1+\frac{2k(d-3)(d-4)\alpha(\varepsilon) g(\varepsilon )^{2}}{r_{+}^{2}%
}\right] },  \label{T+} \\
\eta _{+} &=&\left. \eta \right\vert _{r=r_{+}},  \notag
\end{eqnarray}%
which shows that rainbow functions, the nonlinearity and GB parameters
affect the black hole temperature.

Now, we are going to calculate entropy. Regarding Einstein gravity, it was
shown that the entropy of black holes satisfies the so-called area law which
states the black hole entropy is equal to one-quarter of horizon area \cite%
{Bekenstein1A,Bekenstein1B,Bekenstein1C,Hawking1A,Hawking1B}. However, we
could not use the area law for higher curvature gravity \cite{fails1,fails2}%
. It is known that the entropy in of asymptotically flat solutions ($%
\Lambda(\varepsilon) =0$) of GB gravity can be obtained from Wald formula
\cite{Wald1,Wald2,Wald3,Wald4,Wald5}
\begin{equation}
S=\frac{1}{4}\int d^{d-2}x\sqrt{\tilde{g}}\left( 1+2\alpha(\varepsilon)
\widetilde{R}\right) =\frac{r_{+}^{d-2}}{4g(\varepsilon )^{d-2}}\left( 1+%
\frac{2k\left( d-2\right) (d-3)\alpha(\varepsilon) g(\varepsilon )^{2}}{%
r_{+}^{2}}\right) ,  \label{ents}
\end{equation}%
where the integration is done on the $(d-2)$-dimensional spacelike
hypersurface with the induced metric $\tilde{g}_{\mu \nu }$. We should note
that $\tilde{g}$ is the determinant of $\tilde{g}_{\mu \nu }$ and $\tilde{R}$
is the Ricci scalar for the induced metric. Following Gibbs-Duhem relation,
one can show that the entropy of asymptotically (a)dS black holes in GB
gravity is the same as Eq. (\ref{ents}). It is notable that although the
(nonlinear) electromagnetic source changes the locations of inner and outer
horizons, it does not change the functional form of entropy formula and also
area law (see Ref. \cite{entropyCharge} for more details).

Calculating the flux of the electric field through a given closed
hypersurface at infinity, one can obtain the electric charge per unit volume
$V_{d-2}$ of the black hole, yielding%
\begin{equation}
Q=\frac{q(\varepsilon) f(\varepsilon )}{4\pi g(\varepsilon )^{d-3}},
\label{charge}
\end{equation}%
which shows that although the total charge does not depend on the
nonlinearity of the electromagnetic field, it depends on the
energy functions. This behavior is expected, since for large
values of radial coordinate, the electric field of BI theory
reduces to linear Maxwell field, but metric components depend on
the energy functions. Regarding $\partial _{t}$ as the Killing
vector ($\chi ^{\mu }$) of static spacetime, we can obtain the
electric potential of event horizon with respect to the potential
reference
\begin{equation}
\Phi =A_{\mu }\chi ^{\mu }\left\vert _{r\rightarrow \infty }-A_{\mu }\chi
^{\mu }\right\vert _{r=r_{+}}=\frac{q(\varepsilon) \mathcal{H}_{+}}{%
(d-3)r_{+}^{d-3}},  \label{potential}
\end{equation}%
where $\mathcal{H}_{+}=\mathcal{H}(r=r_{+})$.

Considering the behavior of the metric at large $r$, one can obtain the ADM
(Arnowitt-Deser-Misner) mass of black hole for asymptotically flat solutions
\cite{Brewin}. In addition, we use the counterterm method to calculate the
finite mass of (a)dS solutions. Both calculations (ADM and counterterm
methods) lead to the following unique relation for the mass per unit volume $%
V_{d-2}$
\begin{equation}
M=\frac{(d-2)m(\varepsilon)}{16\pi f(\varepsilon )g(\varepsilon )^{d-1}}.
\label{mass}
\end{equation}

Since all conserved and thermodynamic quantities have been obtained, we can
examine the validity of the first law of thermodynamics. Considering the
expressions for the entropy, the electric charge and the mass given in Eqs. (%
\ref{ents}), (\ref{charge}) and (\ref{mass}), we can obtain the total mass $%
M $\ as a function of the extensive quantities $Q$ and $S$, such as
Smarr-type formula
\begin{equation}
M(S,Q)=\frac{f(\varepsilon )^{2}r_{+}^{d-1}}{16(d-1)}\left( \frac{%
4(d-2)\Omega \beta(\varepsilon) ^{2}\mathcal{A}}{(d-3)}+4\beta(\varepsilon)
^{2}(1-\sqrt{1+\Omega })-2\Lambda(\varepsilon) +\Sigma \right) ,
\label{Smark1}
\end{equation}%
where $r_{+}=r_{+}(S)$ and
\begin{eqnarray*}
\Sigma &=&\frac{(d-1)(d-2)kg(\varepsilon )^{2}}{r_{+}^{2}}\left[ 1+\frac{%
(d-3)(d-4)k\alpha(\varepsilon) g(\varepsilon )^{2}}{r_{+}^{2}}\right] , \\
\mathcal{A} &=&{}_{2}F_{1}\left( \left[ \frac{1}{2},\frac{d-3}{2d-4}\right] ,%
\left[ \frac{3d-7}{2d-4}\right] ,-\Omega \right) , \\
\Omega &=&\frac{16\pi ^{2}g(\varepsilon )^{2d-4}Q^{2}}{\beta(\varepsilon)
^{2}f(\varepsilon )^{2}r_{+}^{2d-4}}.
\end{eqnarray*}

Now, we regard $Q$ and $S$ as the extensive parameters and introduce their
conjugate intensive parameters which are, respectively, the temperature and
the electric potential, as
\begin{eqnarray}
&&T=\left( \frac{\partial M}{\partial S}\right) _{Q}=\left( \frac{\partial M%
}{\partial r_{+}}\right) _{Q}\left/ \left( \frac{\partial S}{\partial r_{+}}%
\right) _{Q}\right. ,  \label{Tk1} \\
&&\Phi =\left( \frac{\partial M}{\partial Q}\right) _{S}=\left( \frac{%
\partial M}{\partial q}\right) _{S}\left/ \left( \frac{\partial Q}{\partial q%
}\right) _{S}\right. .  \label{Phik1}
\end{eqnarray}

It is straightforward to show that Eqs. (\ref{Tk1}) and
(\ref{Phik1}) are equal to Eqs. (\ref{T+}) and (\ref{potential}),
respectively, and therefore, one can conclude that these
quantities satisfy the first law of thermodynamics
\begin{equation}
dM=TdS+\Phi dQ.  \label{FirstLawk1}
\end{equation}

Now, we are going to check thermal stability. Thermodynamic
stability of black holes can be investigated in various ensembles.
Referring to thermal stability criterion in the canonical
ensemble, stable black holes have positive heat capacity
($C_{Q}=T_{+}/\left( \partial ^{2}M/\partial S^{2}\right) _{Q}$ ).
As an additional note, we should mention that since size plays an
essential role in gravitational thermodynamics, the negativity of
the heat capacity does no, necessarily, mean that a system is
unstable in the canonical ensemble. There are some finite size
black hole systems with negative heat capacity but stable state
\cite{York}. Since the black hole systems, here, have infinite
size, negative $C_{Q}$ indicates unstable state.

The canonical ensemble instability criterion is sufficiently strong to veto
some gravity toy models. Thermal stability of a typical thermodynamic system
should be considered as the result of small variations of the thermodynamic
coordinates. In this regard, the energy $\mathcal{M}(S,Q)$ should be a
convex function of its extensive variable. The electric charge is a fixed
parameter in the canonical ensemble and the positivity of heat capacity is
sufficient to ensure (local) thermal stability. After some calculations, we
find
\begin{equation}
C_{Q}=\frac{(d-2)r_{+}^{d}\left[ 1+\frac{2(d-3)(d-4)k\alpha(\varepsilon)
g(\varepsilon )^{2}}{r_{+}^{2}}\right] ^{2}\left[ 2\Lambda(\varepsilon)
-4\beta(\varepsilon) ^{2}(1-\sqrt{1+\eta _{+}})-A_{0}\right] }{%
4(d-3)g(\varepsilon )^{d}\left[ 2(d-4)k\alpha(\varepsilon) A_{1}+(d-2)kA_{2}-%
\frac{4q(\varepsilon)^{2}\left( 1+A_{3}\right) }{r_{+}^{2d-6}\sqrt{1+\eta
_{+}}}\right] },  \label{CQ}
\end{equation}%
where
\begin{eqnarray*}
A_{0} &=&\frac{(d-2)(d-3)kg(\varepsilon )^{2}}{r_{+}^{2}}\left[ 1+\frac{%
(d-4)(d-5)\alpha(\varepsilon) kg(\varepsilon )^{2}}{r_{+}^{2}}\right] , \\
&& \\
A_{1} &=&\frac{(d-2)(d-3)kg(\varepsilon )^{2}}{r_{+}^{2}}\left( \frac{%
(d-4)(d-5)k\alpha(\varepsilon) g(\varepsilon )^{2}}{r_{+}^{2}}-1\right)
-12\left(\beta(\varepsilon) ^{2}-\frac{\Lambda(\varepsilon) }{2}\right), \\
&& \\
A_{2} &=&1+\frac{3(d-4)(d-5)k\alpha(\varepsilon) g(\varepsilon )^{2}}{%
r_{+}^{2}}-\frac{4r_{+}^{2} }{(d-2)(d-3)kg(\varepsilon )^{2}}
\left(\beta(\varepsilon) ^{2}-\frac{\Lambda(\varepsilon) }{2}\right), \\
&& \\
A_{3} &=&\frac{2(d-4)(d-5)k\alpha(\varepsilon) g(\varepsilon )^{2}}{r_{+}^{2}%
}-\frac{\beta(\varepsilon) ^{2}r_{+}^{2d-4}}{q(\varepsilon)^{2}}\left( \frac{%
1}{(d-3)g(\varepsilon )^{2}}+\frac{6(d-4)k\alpha(\varepsilon) }{r_{+}^{2}}%
\right) .
\end{eqnarray*}

\begin{figure}[tbp]
$%
\begin{array}{cc}
\epsfxsize=7cm \epsffile{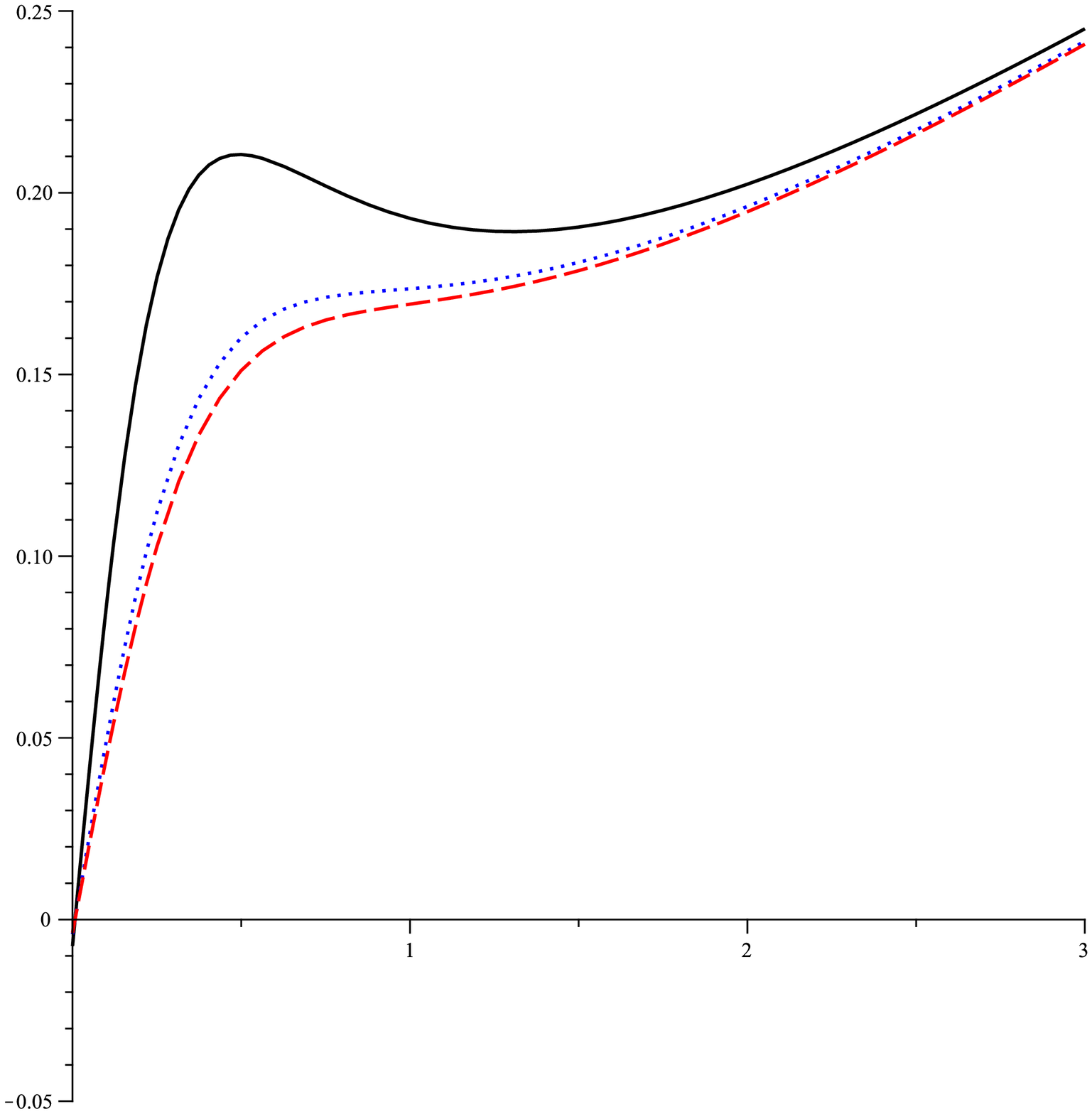} & \epsfxsize=7cm \epsffile{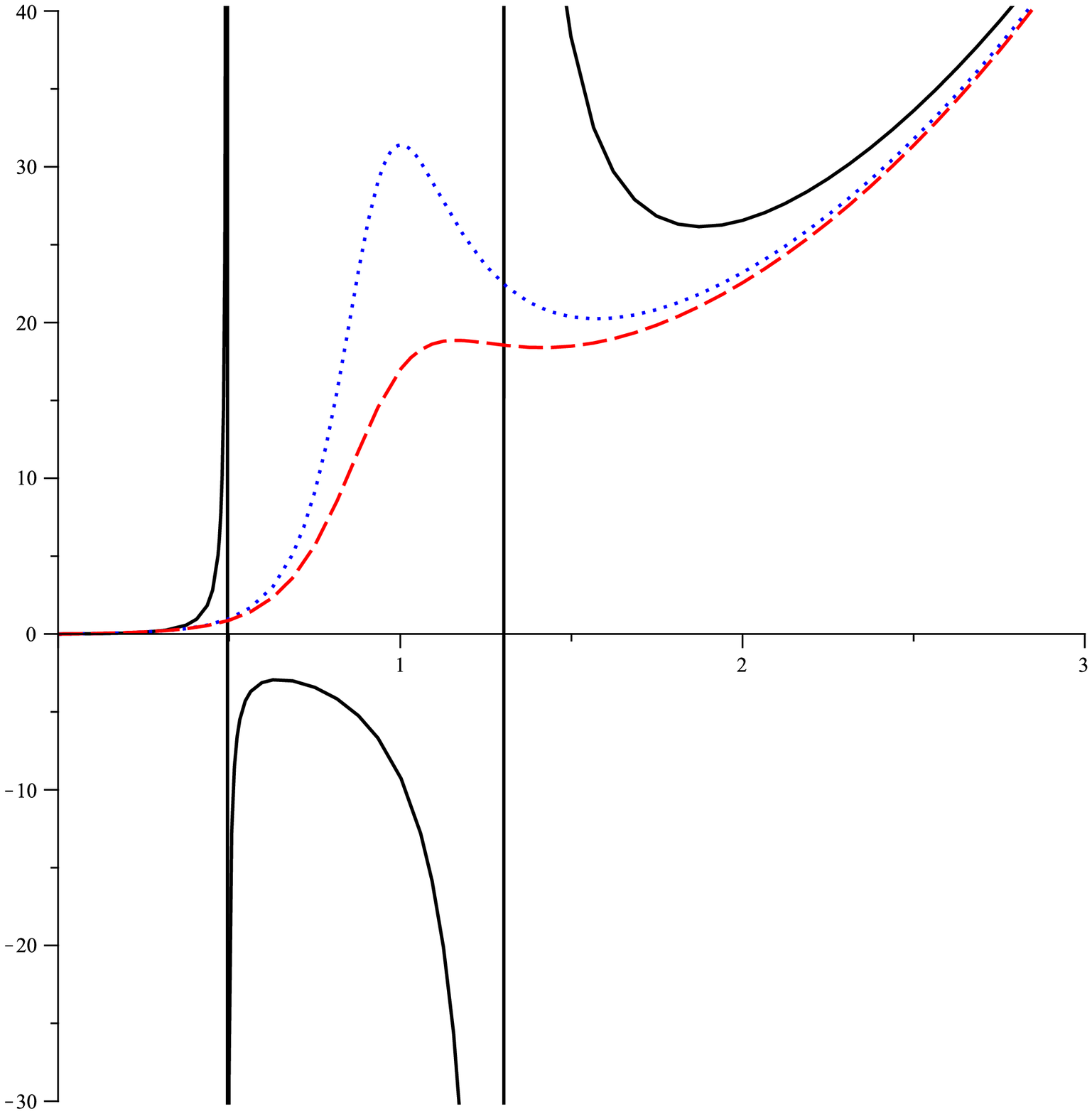}%
\end{array}
$%
\caption{$T$ (left) and $C_{Q}$ (right) diagrams versus $r_{+}$ for $d=5$, $%
g(\protect\varepsilon)=f(\protect\varepsilon)=0.9$, $\Lambda(\protect%
\varepsilon)=-1$, $\protect\beta(\protect\varepsilon)=0.1$, $q(\protect%
\varepsilon)=0.1$, and $\protect\alpha(\protect\varepsilon)=0.05$
(continuous line), $\protect\alpha(\protect\varepsilon)=0.09$ (dotted line)
and $\protect\alpha(\protect\varepsilon)=0.10$ (dashed line). }
\label{Fig-alpha}
\end{figure}

\begin{figure}[tbp]
$%
\begin{array}{cc}
\epsfxsize=7cm \epsffile{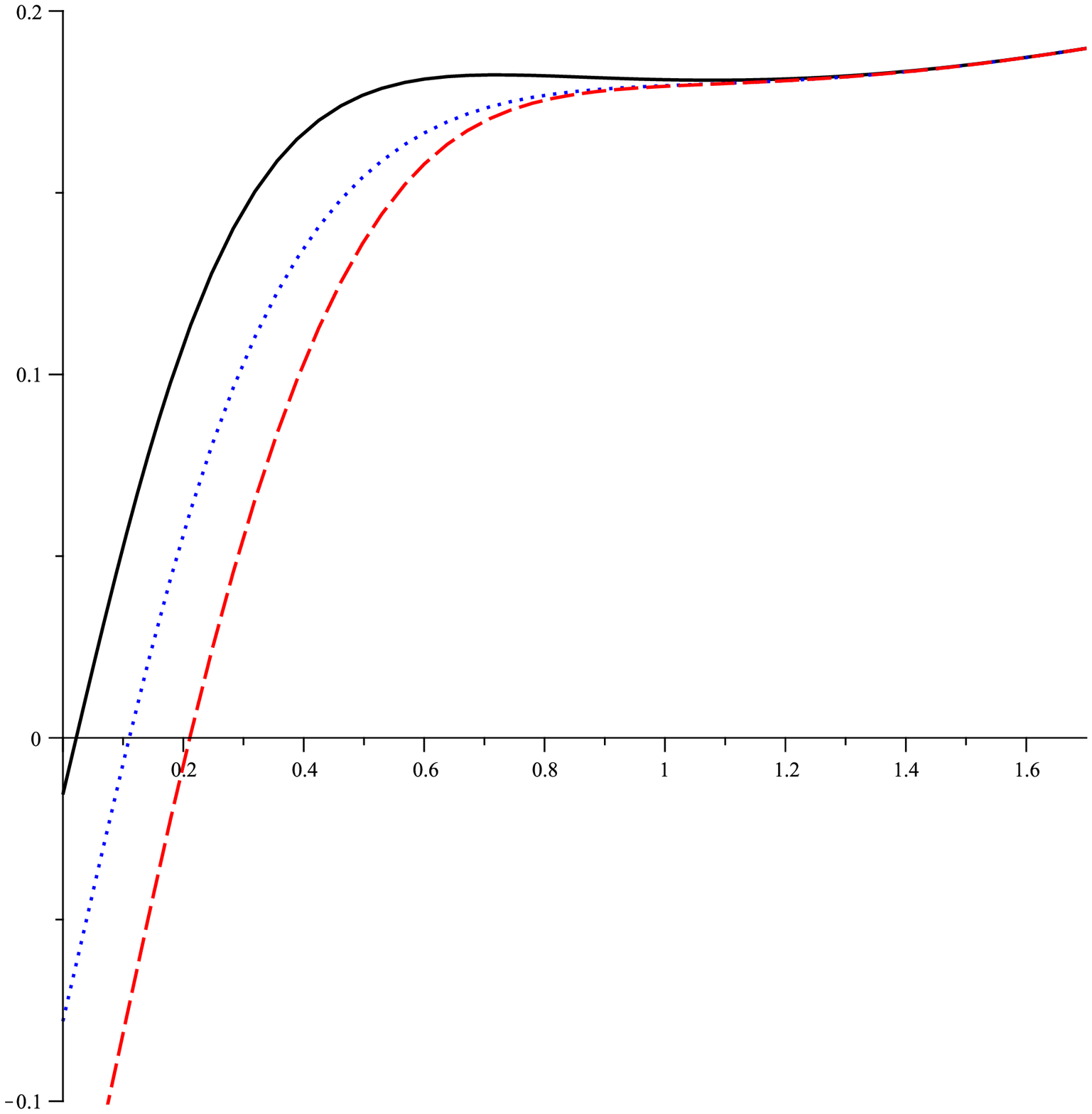} & \epsfxsize=7cm \epsffile{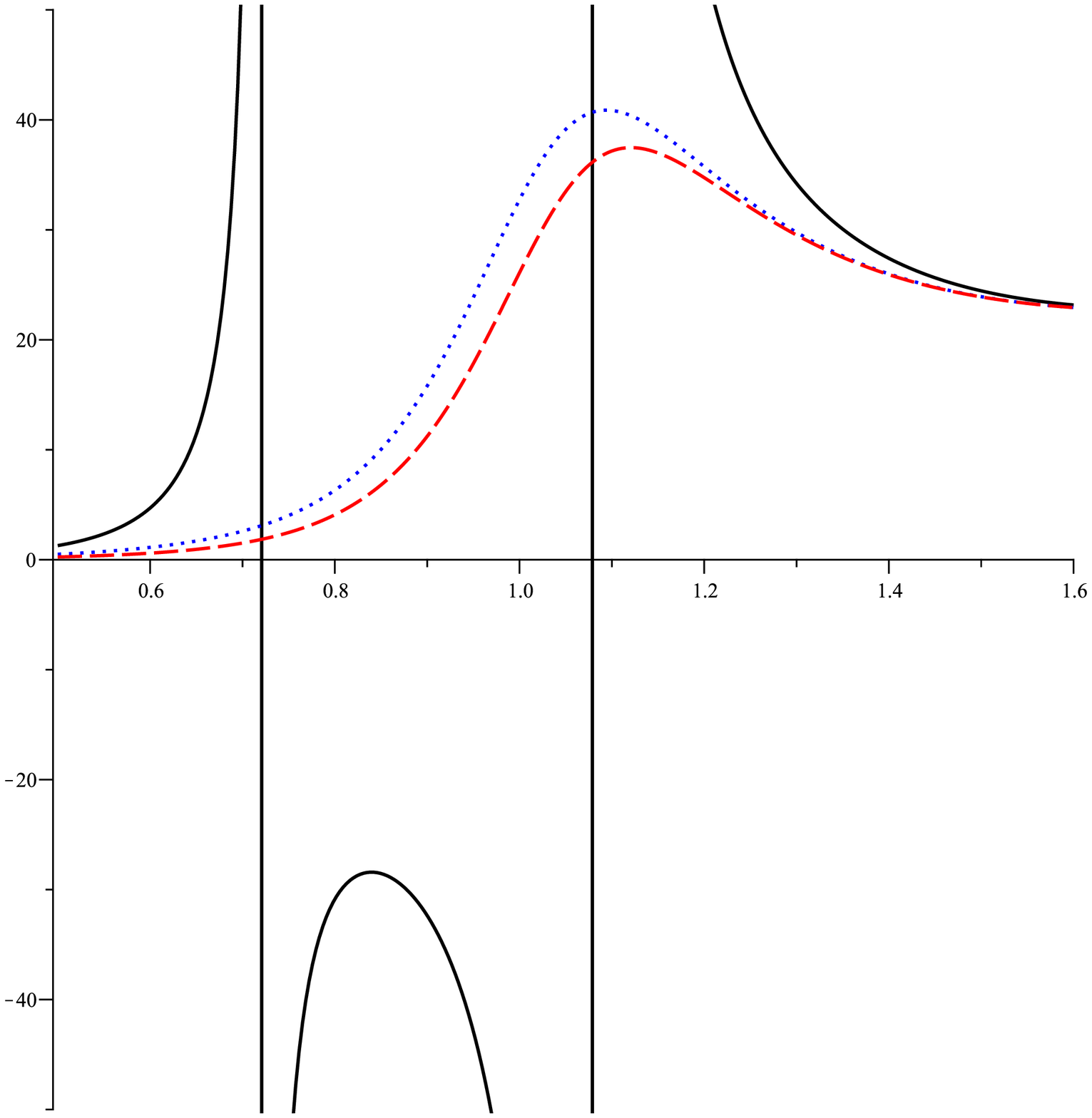}%
\end{array}
$%
\caption{$T$ (left) and $C_{Q}$ (right) diagrams versus $r_{+}$ for $d=5$, $%
g(\protect\varepsilon)=f(\protect\varepsilon)=0.9$, $\Lambda(\protect%
\varepsilon)=-1$, $\protect\alpha(\protect\varepsilon)=0.07$, $q(\protect%
\varepsilon)=0.3$, and $\protect\beta(\protect\varepsilon)=0.1$ (continuous
line), $\protect\beta(\protect\varepsilon)=0.5$ (dotted line) and $\protect%
\beta(\protect\varepsilon)=1.0$ (dashed line).}
\label{Fig-beta}
\end{figure}

\begin{figure}[tbp]
$%
\begin{array}{cc}
\epsfxsize=7cm \epsffile{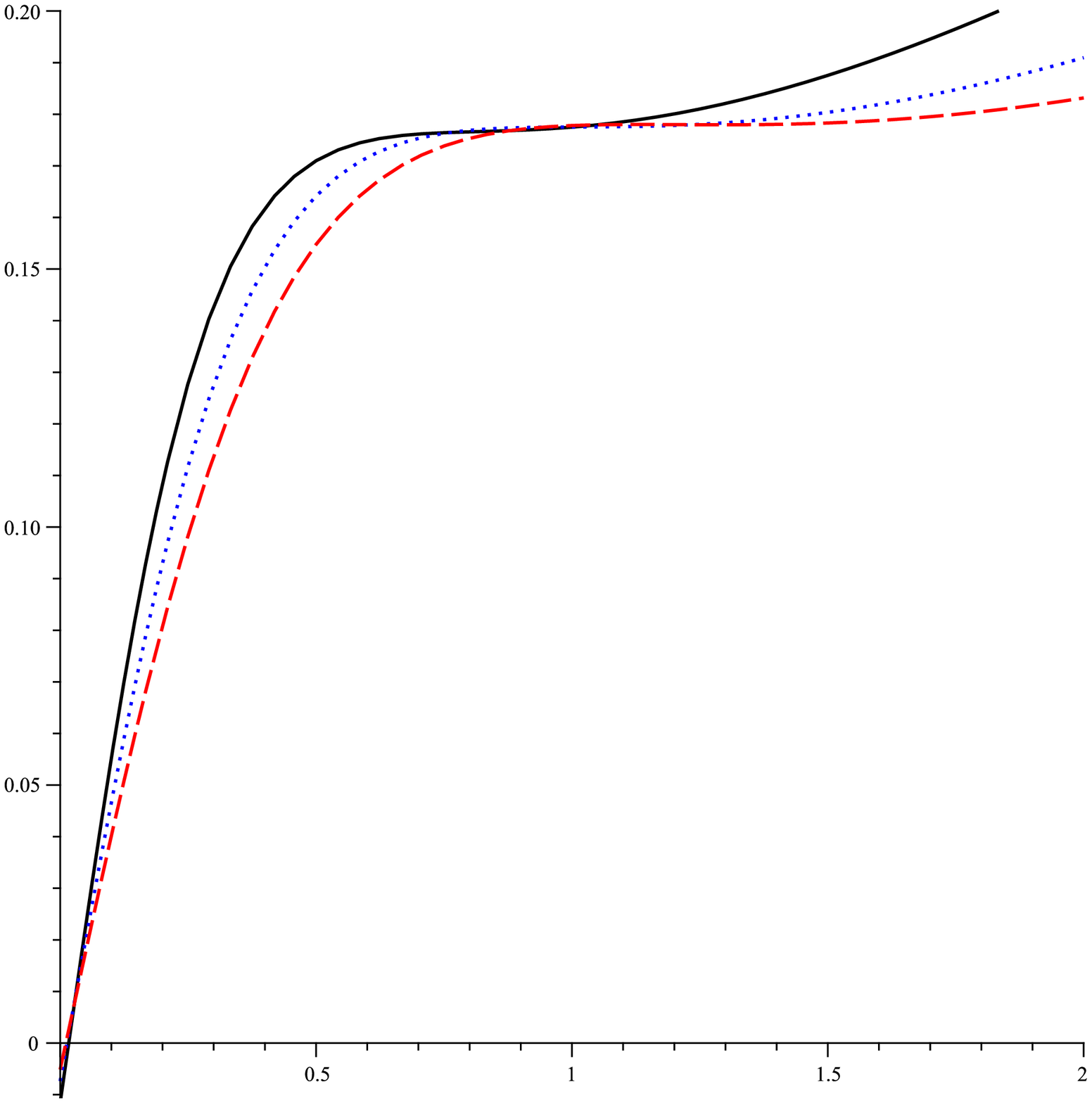} & \epsfxsize=7cm \epsffile{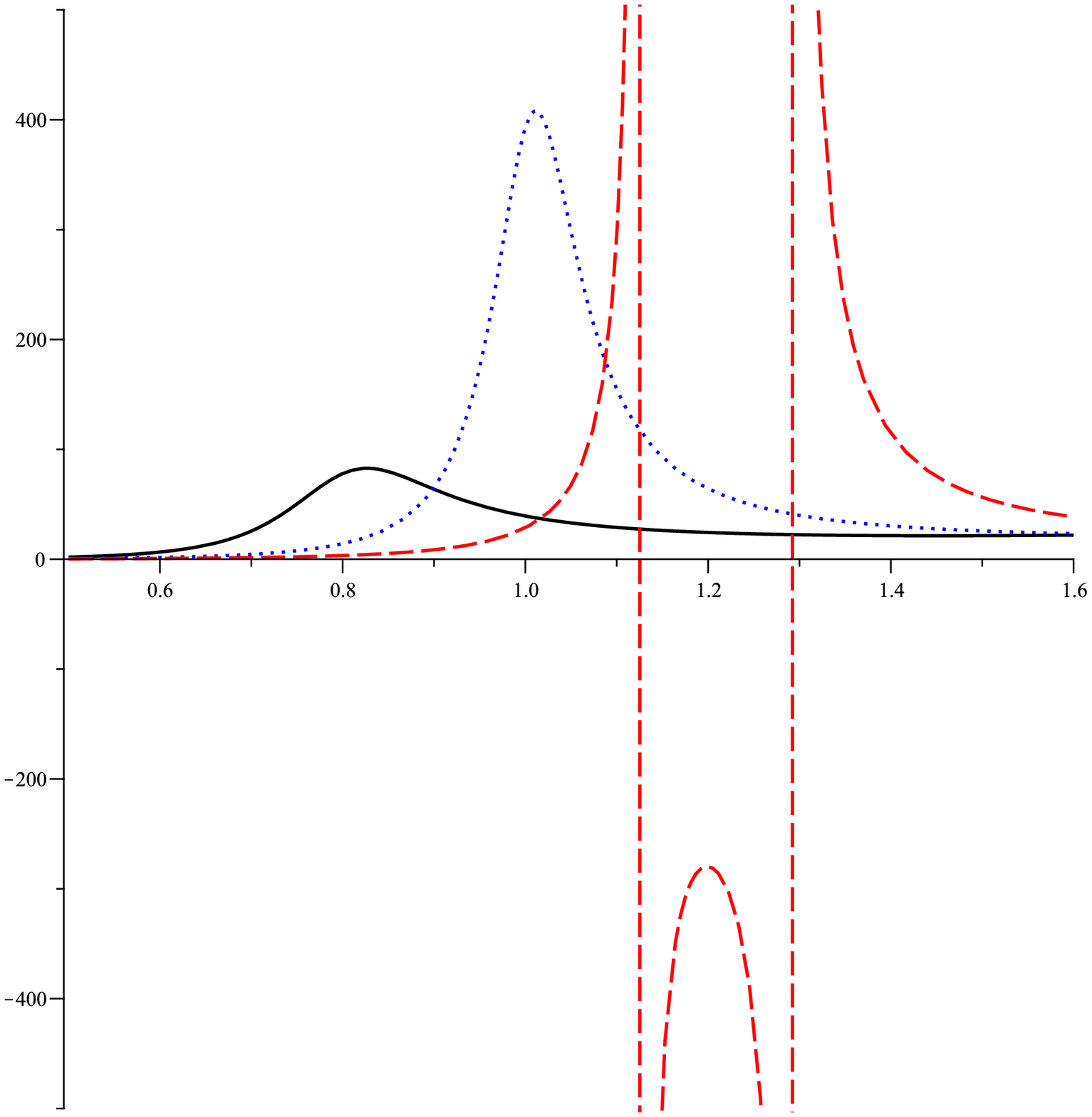}%
\end{array}
$%
\caption{$T$ (left) and $C_{Q}$ (right) diagrams versus $r_{+}$
for $d=5$, $\Lambda(\protect\varepsilon)=-1$, $\protect%
\alpha(\protect\varepsilon)=0.08$,
$\protect\beta(\protect\varepsilon)=0.1$,
$q(\protect\varepsilon)=0.2$, and
$f(\protect\varepsilon)=g(\protect\varepsilon)=0.8$ (continuous
line), $f(\protect\varepsilon)=g(\protect\varepsilon)=1.0$ (dotted line) and $f(\protect\varepsilon)=g(\protect\varepsilon%
)=1.2$ (dashed line).} \label{Fig-f(e)}
\end{figure}

It is a nontrivial task to analytically find the positivity of the
heat capacity. Instead, we use the numerical analysis with various
plots to discuss the sign and divergence points of heat capacity.
Regarding Figs. \ref{Fig-alpha}--\ref{Fig-f(e)} with selected
parameters, one finds that two different cases may take place for
the heat capacity.


In the first case, $C_{Q}$ is positive definite for $r_{+}>r_{0}$, in which $%
r_{0}$ is the real positive root of temperature. Also, variation of free
parameters in this case leads to changing $r_{0}$. In this regard, we study
the effects of GB coefficient ($\alpha(\varepsilon) $), BI parameter ($%
\beta(\varepsilon) $) and rainbow function ($g(\varepsilon )$).
Left panels of Figs. \ref{Fig-alpha}--\ref{Fig-f(e)} and numerical
analysis show that although variation of $\alpha(\varepsilon) $
does not have significant effect on $r_{0}$, and $r_{0}$ is an
increasing function of $\beta(\varepsilon) $ (decreasing function
of $g(\varepsilon )$).

In the second case, there are two divergence points ($r_{+1}$ and $r_{+2}$
with $r_{+1}<r_{+2}$ ) for the heat capacity. In other words, the heat
capacity is positive for $r_{0}<r_{+}<r_{+1}$ and $r_{+}>r_{+2}$ and black
hole is stable in these ranges. However, there is an unstable state for $%
r_{+1}<r_{+}<r_{+2}$. Considering phase space conception, one may consider a
phase transition between the mentioned two stable states. It is worthwhile
to mention that variation of free parameters may change the locations of $%
r_{+1}$ and $r_{+2}$. In addition, right panels of Figs. \ref{Fig-alpha}--%
\ref{Fig-f(e)} indicate that the unstable range of event horizon
is an increasing function of $g(\varepsilon )$ and a decreasing
function of $\alpha(\varepsilon) $ and $%
\beta(\varepsilon) $. In other words, one can find a limit value
for GB coefficient ($\alpha _{0}(\varepsilon)$) BI parameter
($\beta _{0}(\varepsilon)$) and rainbow function
($g_{0}(\varepsilon )$), in which for $\alpha(\varepsilon) >\alpha
_{0}(\varepsilon)$ ($\beta(\varepsilon)
>\beta _{0}(\varepsilon)$ or $g(\varepsilon )<g_{0}(\varepsilon )$), there
is no unstable phase and the heat capacity is positive definite in the range
of $r_{+}>r_{0}$.


\section{Asymptotically adS rotating black branes\label{rotating}}

In this section, we are going to add a global rotation into the zero
curvature horizon ($k=0$). The maximum number of independent rotation
parameters is equal to $[\frac{d-1}{2}]$ ($[x]$ is the integer part of $x$)
which is due to the fact that the rotation group in $d-$dimensions is $%
SO(d-1)$. Now, we perform the following local boost transformation in the $%
t-\phi _{i}$ planes for adding the angular momentum
\begin{equation}
\frac{t}{f(\varepsilon )}\mapsto \Xi(\varepsilon) \frac{t}{f(\varepsilon )}%
-a_{i}(\varepsilon)\frac{\phi _{i}}{g(\varepsilon )},\hspace{0.5cm}\frac{%
\phi _{i}}{g(\varepsilon )}\mapsto \Xi(\varepsilon) \frac{\phi _{i}}{%
g(\varepsilon )}-\frac{a_{i}(\varepsilon)}{l(\varepsilon)^{2}}\frac{t}{%
f(\varepsilon )},  \label{Tr}
\end{equation}%
where $l(\varepsilon)$ is an energy dependent scale parameter with length
dimension, $a_{i}(\varepsilon)$ is rotation parameters in which $i=1...[%
\frac{d-1}{2}]$ and $\Xi(\varepsilon) =\sqrt{1+\sum_{i=1}^{p}\frac{%
a_{i}(\varepsilon)^{2}}{l(\varepsilon)^{2}}}$. Applying such boost into the
metric (\ref{Metric}) with $k=0$, one can obtain the following $d$%
-dimensional Ricci-flat solutions with $p\leq \lbrack \frac{d-1}{2}]$
rotation parameters
\begin{eqnarray}
ds^{2} &=&-\psi (r)\left( \frac{\Xi(\varepsilon) }{f(\varepsilon )}dt-\frac{1%
}{g(\varepsilon )}{{\sum_{i=1}^{p}}}a_{i}(\varepsilon)d\phi _{i}\right) ^{2}+%
\frac{r^{2}}{l^{4}}{{\sum_{i=1}^{p}}}\left( \frac{a_{i}(\varepsilon)}{%
f(\varepsilon )}dt-\frac{\Xi(\varepsilon) l(\varepsilon)^{2}}{g(\varepsilon )%
}d\phi _{i}\right) ^{2}  \notag \\
&&\ \text{ }+\frac{dr^{2}}{g(\varepsilon )^{2}\psi (r)}-\frac{r^{2}}{%
g(\varepsilon )^{2}l(\varepsilon)^{2}}{\sum_{i<j}^{p}}[a_{i}(\varepsilon)
d\phi _{j}-a_{j}(\varepsilon)d\phi _{i}]^{2}+\frac{r^{2}}{g(\varepsilon )^{2}%
}{{\sum_{i=p+1}^{d-2}}}d\phi _{i}^{2}.  \label{Metric2}
\end{eqnarray}

Using Eq. (\ref{Maxeq}), we can obtain the following consistent gauge
potential
\begin{equation}
A_{\mu }=h(r)\left( \frac{\Xi(\varepsilon) }{f(\varepsilon )}\delta _{\mu
}^{0}-\frac{a_{i}(\varepsilon)}{g(\varepsilon )}\delta _{\mu }^{i}\right)
\text{ \ \ (no sum on }i\text{)}  \label{Amu2}
\end{equation}%
where $h(r)$ is the same as Eq. (\ref{h(r)}). In addition, straightforward
calculations show that all components of the gravitational field equation (%
\ref{Geq}) can be satisfied by the metric function of Eq. (\ref{fr}). It is
worthwhile to mention that the local transformation (\ref{Tr}) generates a
new metric, because it is not a proper coordinate transformation on the
entire manifold \cite{Stachel}. In other words, the static and rotating
metrics can be locally mapped into each other but not globally, and
therefore they are distinct.

Following the same method, one can find that there is a curvature
singularity at $r=0$, which can be covered with an event horizon. Therefore,
one may interpret such singularity as a black brane. Using the fact that $%
\chi =\partial _{t}+{\sum_{i}}\Omega _{i} \partial _{\phi _{i}}$ is the null
generator of the horizon for the mentioned rotating black branes, we can
obtain%
\begin{equation}
{T}_{rotating}{=}\left. \frac{{T}_{static}}{\Xi(\varepsilon) }\right\vert
_{k=0},  \label{Tem}
\end{equation}%
\begin{equation}
\Phi _{rotating}{=}\frac{\Phi _{static}}{\Xi(\varepsilon) },  \label{Pot}
\end{equation}%
\begin{equation}
Q_{rotating}=\Xi(\varepsilon) Q_{static},  \label{charge2}
\end{equation}%
\begin{equation}
S_{rotating}=\left. \Xi(\varepsilon) S_{static}\right\vert _{k=0},
\label{ent2}
\end{equation}%
\begin{equation}
M_{rotating}=\left( \frac{(d-1)\Xi(\varepsilon) ^{2}-1}{d-2}\right)
M_{static},  \label{mass2}
\end{equation}%
and for rotating quantities, one can, respectively, obtain the following
functional forms of angular velocities and angular momenta
\begin{equation}
\Omega _{i}=\frac{a_{i}(\varepsilon)}{\Xi(\varepsilon) l(\varepsilon)^{2}},
\label{Om}
\end{equation}%
\begin{equation}
J_{i}=\frac{d-1}{16\pi }\Xi(\varepsilon) m(\varepsilon) a_{i}(\varepsilon).
\label{AngMom}
\end{equation}

In order to examine the validity of the first law of thermodynamics, we
regard $S$, $Q$ and $J_{i}$'s as a complete set of extensive quantities for
the mass $M(S,Q,J_{i})$ and define intensive quantities conjugate to them.
The conjugate quantities are $T$, $\Phi $ and $\Omega _{i}$'s
\begin{eqnarray}
T &=&\left( \frac{\partial \mathcal{M}}{\partial S}\right) _{Q,J_{i}},
\label{Dsmar1} \\
\Phi &=&\left( \frac{\partial \mathcal{M}}{\partial Q}\right) _{S,J_{i}},
\label{Dsmar2} \\
\Omega _{i} &=&\left( \frac{\partial \mathcal{M}}{\partial J_{i}}\right)
_{S,Q}.  \label{Dsmar3}
\end{eqnarray}

Considering the fact that metric function vanishes at the event horizon with
Eqs. (\ref{Dsmar1}), (\ref{Dsmar2}) and (\ref{Dsmar3}), we find that the
intensive quantities calculated by Eqs. (\ref{Dsmar1}), (\ref{Dsmar2}) and (%
\ref{Dsmar3}) are, respectively, consistent with Eqs. (\ref{Tem}), (\ref{Pot}%
) and (\ref{Om}), and therefore, one can confirm that the relevant
thermodynamic quantities satisfy the first law of thermodynamics as
\begin{equation}
dM=TdS+\Phi dQ+\sum_{i}\Omega _{i}dJ_{i}.  \label{FirstLaw}
\end{equation}

Since we investigate thermodynamic stability of adS Ricci-flat
static black holes in previous section. Here, we are going to
check the effect of rotation on thermal stability. For rotating
solutions, the mass $M$ is a function of the entropy $S$, the
angular momentum $J$ and the electric
charge $Q$. As we mentioned before, the positivity of the heat capacity $%
C_{Q,J}=T\left/ \left( \partial ^{2}M/\partial S^{2}\right) _{Q,J}\right. $
is sufficient to ensure the thermodynamic stability. Numerical calculations
show that the effects of variation of rainbow functions, BI parameter and GB
coefficient are the same as those in static case. In addition, as it was
shown in Ref. \cite{GBBI2}, $C_{Q,J}$ is an increasing regular function of $%
\Xi$. Thus we ignore more explanations and in the next section, we focus on
more interesting case $k=1$ for investigating critical behavior.

\section{Critical behavior}

Here, we use the the extended phase space thermodynamics to investigate
critical behavior. To do so, we take into account the adS black holes with
spherical horizons (it was shown that there is not any van der Waals like phase transition for $%
k=0$ and $k=-1$). In this regard, we treat the cosmological
constant as a thermodynamics pressure. In fact, we do not work in
a fixed AdS background, and therefore, the cosmological constant
is not a constant anymore, but a variable. If we treat the
negative cosmological constant proportional to thermodynamic
pressure, its conjugate quantity will be thermodynamic volume.
Using the scaling argument, it has already been shown that the
Smarr relation is consistent with first law of thermodynamics by
assuming the cosmological constant as a thermodynamic variable.
Here, we regard the following relation
between the cosmological constant and the pressure \cite%
{PressureLambda1,PressureLambda2,PressureLambda3,PressureLambda4,
PressureLambda5,PressureLambda6,PressureLambda7,PressureLambda8,
PressureLambda9,PressureLambda10,PressureLambda11})
\begin{equation}
P=-\frac{\Lambda(\varepsilon) }{{8\pi }}.  \label{Pressure}
\end{equation}

As a notable comment, it is worthwhile to mention that it was shown that the
modified Smarr relation which can be calculated by the scaling argument is
completely in agreement with the modified first law of thermodynamics in the
extended phase space. In this regard, in addition to $\Lambda(\varepsilon) $%
, the nonlinearity parameter ($\beta(\varepsilon) $) and GB parameter ($%
\alpha(\varepsilon)$) are two other thermodynamical variables \cite%
{Smarr1,Smarr2}. It is notable that in the presence of rainbow functions,
the modified first law of thermodynamics can be written as
\begin{equation}
dM=TdS+\Phi dQ+VdP+\mathcal{A}d\alpha +\mathcal{B}d\beta ,
\label{GenFirstLaw}
\end{equation}%
where the conjugate quantities can be obtained as
\begin{eqnarray*}
T &=&\left( \frac{\partial M}{\partial S}\right) _{Q,P,\alpha ,\beta }, \\
\Phi &=&\left( \frac{\partial M}{\partial Q}\right) _{S,P,\alpha ,\beta }, \\
V &=&\left( \frac{\partial M}{\partial P}\right) _{S,Q,\alpha ,\beta }, \\
\mathcal{A} &=&\left( \frac{\partial M}{\partial \alpha }\right)
_{S,Q,P,\beta }, \\
\mathcal{B} &=&\left( \frac{\partial M}{\partial \beta }\right)
_{S,Q,P,\alpha }.
\end{eqnarray*}

Moreover, we can obtain the generalized Smarr relation for our
asymptotically adS solutions in the extended phase space
\begin{equation}
(d-3)M=(d-2)TS+(d-3)Q\Phi -2PV-2\left( \frac{d-2}{d-4}\right) \mathcal{A}%
\alpha -\mathcal{B}\beta ,  \label{Smarr2}
\end{equation}%
which is completely in agreement with the mentioned modified first law of
thermodynamics.

Inserting Eq. (\ref{Pressure}) in the temperature equation (\ref{T+}), one
can find equation of state $P=P(T,r_{+})$
\begin{eqnarray}
P &=&\frac{4\beta(\varepsilon) ^{2}}{16\pi }\left( \sqrt{1+\eta _{+}}%
-1\right) +\left( 1+\frac{2(d-3)(d-4)\alpha(\varepsilon) g(\varepsilon )^{2}%
}{r_{+}^{2}}\right) \frac{(d-2)f(\varepsilon )g(\varepsilon )T}{4r_{+}}-
\notag \\
&&\frac{(d-2)(d-3)g(\varepsilon )^{2}}{16\pi r_{+}^{2}}\left( 1+\frac{%
(d-4)(d-5)\alpha(\varepsilon) g(\varepsilon )^{2}}{r_{+}^{2}}\right) .
\label{EOS}
\end{eqnarray}

Since the critical point is an inflection point on the critical isothermal $%
P-r_{+}$ diagrams, we use the following relations to obtain the proper
equations for critical quantities
\begin{equation}
\left( \frac{\partial P}{\partial r_{+}}\right) _{T}=0,  \label{inflection1}
\end{equation}
\begin{equation}
\left( \frac{\partial ^{2}P}{\partial r_{+}^{2}}\right) _{T}=0.
\label{inflection2}
\end{equation}

Using Eqs. (\ref{inflection1}) and (\ref{inflection2}), one can find the
critical quantities. Numerical calculations confirm that by choosing
suitable parameters, one can find critical values for thermodynamic
quantities. In order to analyze the effects of various parameters, we
present different tables (see tables $1-6$). Accordingly, we can find that
GB and nonlinearity parameters, rainbow functions and dimensionality affect
the critical quantities. Based on tables $1$ and $2$, one finds increasing
GB parameter leads to increasing critical volume and decreasing critical
temperature and pressure. In addition, for sufficiently large $%
\alpha(\varepsilon)$, one can obtain the universality ratio $1/4 $
independent of other parameters. It is notable that if we increase $%
\alpha(\varepsilon) $ more, we cannot obtain a real positive
values for the critical quantities. In other words, there is an
$\alpha _{0}(\varepsilon)$ in which for $\alpha(\varepsilon)
>\alpha _{0}(\varepsilon)$ there is no phase transition and
critical behavior (see Fig. \ref{Fig-alpha} for more details). We
should note that the value of $\alpha _{0}(\varepsilon)$ depends
on other parameters. The same behavior takes place for the
nonlinearity parameter. It means that critical volume is (critical
temperature and pressure are) an increasing function (decreasing
functions)
of BI parameter, and also there is a $\beta _{0}(\varepsilon)$ in which for $%
\beta(\varepsilon) >\beta _{0}(\varepsilon)$ system is thermally
stable and there is no phase transition (see Fig. \ref{Fig-beta}
for more details). For the rainbow functions, we obtain different
behavior. We find that by increasing these functions, all critical
quantities are increasing, except critical temperature. In
addition, there is a lower bound for the rainbow functions, in
which if we regard the values of the rainbow functions
less than such bound, we cannot find any phase transition (see Fig. \ref%
{Fig-f(e)} for more details).\newline



\begin{center}
\begin{tabular}{|c|c|c|}
\hline
\begin{tabular}{ccccc}
\hline\hline
\hspace{0.4cm}$\alpha(\varepsilon) $\hspace{0.4cm} & \hspace{0.4cm}$r_{c}$%
\hspace{0.4cm} & \hspace{0.4cm}$T_{c}$\hspace{0.4cm} & \hspace{0.4cm}$P_{c}$%
\hspace{0.4cm} & \hspace{0.4cm}$\frac{P_{c}r_{c}}{T_{c}}$ \\ \hline\hline
$0.0100$ & 0.4606 & 0.3456 & 0.1868 & 0.2490 \\ \hline
$0.0500$ & 0.8822 & 0.1807 & 0.0505 & 0.2468 \\ \hline
$0.1000$ & 1.2072 & 0.1328 & 0.0270 & 0.2458 \\ \hline
$0.5000$ & 2.6086 & 0.0636 & 0.0062 & 0.2561 \\ \hline
$1.0000$ & 3.5479 & 0.0456 & 0.0032 & 0.2528 \\ \hline
$10.0000$ & 10.9575 & 0.0145 & 0.0003 & 0.2500 \\ \hline
\end{tabular}
& \ \ \ \ \ \  &
\begin{tabular}{ccccc}
\hline\hline
\hspace{0.4cm}$\alpha(\varepsilon) $\hspace{0.4cm} & \hspace{0.4cm}$r_{c}$
\hspace{0.4cm} & \hspace{0.4cm} $T_{c}$\hspace{0.4cm} & \hspace{0.4cm} $P_{c}
$\hspace{0.4cm} & \hspace{0.4cm}$\frac{P_{c}r_{c}}{T_{c}}$ \\ \hline\hline
$0.01000$ & 0.6790 & 0.3720 & 0.1869 & 0.3411 \\ \hline
$0.05000$ & 1.1013 & 0.1938 & 0.0547 & 0.3106 \\ \hline
$0.10000$ & 1.4543 & 0.1410 & 0.0295 & 0.3048 \\ \hline
$0.50000$ & 2.0513 & 0.0647 & 0.0064 & 0.2041 \\ \hline
$1.00000$ & 3.2858 & 0.0459 & 0.0033 & 0.2367 \\ \hline
$10.0000$ & 10.9384 & 0.0145 & 0.0003 & 0.2496 \\ \hline
\end{tabular}
\\ \hline
\end{tabular}
\\[0pt]
Table $1$ (left): Critical values for $q(\varepsilon)=1$, $f(\varepsilon
)=g(\varepsilon )=1$, $\beta(\varepsilon) =0.1$ and $d=5$.\\[0pt]
Table $2$ (right): Critical values for $q(\varepsilon)=1$, $f(\varepsilon
)=g(\varepsilon )=1$, $\beta(\varepsilon) =0.1$ and $d=6$.
\end{center}


\begin{center}
\begin{tabular}{|c|c|c|}
\hline
\begin{tabular}{ccccc}
\hline\hline
\hspace{0.6cm}$\beta(\varepsilon) $ \hspace{0.6cm} & \hspace{0.4cm}$r_{c}$%
\hspace{0.4cm} & \hspace{0.4cm}$T_{c}$\hspace{0.4cm} & \hspace{0.4cm}$P_{c}$%
\hspace{0.4cm} & \hspace{0.4cm}$\frac{P_{c}r_{c}}{T_{c}}$ \\ \hline\hline
$0.0100$ & 0.3566 & 0.4464 & 0.3130 & 0.2500 \\ \hline
$0.0500$ & 0.4000 & 0.3979 & 0.2485 & 0.2498 \\ \hline
$0.1000$ & 0.4606 & 0.3456 & 0.1868 & 0.2490 \\ \hline
$0.5000$ & 1.3362 & 0.1696 & 0.0365 & 0.2877 \\ \hline
$1.0000$ & 1.4922 & 0.1633 & 0.0335 & 0.3059 \\ \hline
$10.0000$ & 1.5265 & 0.1618 & 0.0328 & 0.3093 \\ \hline
\end{tabular}
& \ \ \ \ \ \  &
\begin{tabular}{ccccc}
\hline\hline
\hspace{0.6cm}$\beta(\varepsilon) $ \hspace{0.6cm} & \hspace{0.4cm}$r_{c}$%
\hspace{0.4cm} & \hspace{0.4cm}$T_{c}$ \hspace{0.4cm} & \hspace{0.4cm}$P_{c}$%
\hspace{0.4cm} & \hspace{0.4cm}$\frac{P_{c}r_{c}}{T_{c}}$ \\ \hline\hline
$0.0100$ & 0.2449 & 0.4332 & 0.2210 & 0.1250 \\ \hline
$0.0500$ & 0.5769 & 0.4055 & 0.2317 & 0.3296 \\ \hline
$0.1000$ & 0.6790 & 0.3720 & 0.1869 & 0.3411 \\ \hline
$0.5000$ & 1.2425 & 0.2764 & 0.0892 & 0.4010 \\ \hline
$1.0000$ & 1.3299 & 0.2708 & 0.0850 & 0.4170 \\ \hline
$10.0000$ & 1.3515 & 0.2693 & 0.0838 & 0.4205 \\ \hline
\end{tabular}
\\ \hline
\end{tabular}
\\[0pt]
Table $3$ (left): Critical values for $q(\varepsilon)=1$, $f(\varepsilon
)=g(\varepsilon )=1$, $\alpha(\varepsilon) =10^{-2}$ and $d=5$.\\[0pt]
Table $4$ (right): Critical values for $q(\varepsilon)=1$, $f(\varepsilon
)=g(\varepsilon )=1$, $\alpha(\varepsilon) =10^{-2}$ and $d=6$.
\end{center}



\begin{center}
\begin{tabular}{|c|c|c|}
\hline
\begin{tabular}{ccccc}
\hline\hline
$f(\varepsilon )=g(\varepsilon )$ & \hspace{0.2cm}$r_{c}$\hspace{0.4cm} &
\hspace{0.4cm}$T_{c}$\hspace{0.4cm} & \hspace{0.4cm}$P_{c}$\hspace{0.4cm} &
\hspace{0.4cm}$\frac{P_{c}r_{c}}{T_{c}}$ \\ \hline\hline
$0.9000$ & 1.1092 & 0.1444 & 0.0259 & 0.1986 \\ \hline
$0.9500$ & 1.1578 & 0.1384 & 0.0265 & 0.2216 \\ \hline
$1.0000$ & 1.2072 & 0.1328 & 0.0270 & 0.2458 \\ \hline
$1.0500$ & 1.2573 & 0.1276 & 0.0275 & 0.2714 \\ \hline
$1.1000$ & 1.3079 & 0.1227 & 0.0280 & 0.2981 \\ \hline
\end{tabular}
& \ \ \ \ \ \  &
\begin{tabular}{ccccc}
\hline\hline
$f(\varepsilon )=g(\varepsilon )$ & \hspace{0.2cm}$r_{c}$\hspace{0.4cm} &
\hspace{0.4cm}$T_{c}$\hspace{0.4cm} & \hspace{0.4cm}$P_{c}$\hspace{0.4cm} &
\hspace{0.4cm}$\frac{P_{c}r_{c}}{T_{c}}$ \\ \hline\hline
$0.9000$ & 0.6452 & 0.1480 & 0.0147 & 0.0641 \\ \hline
$0.9500$ & 1.4042 & 0.1477 & 0.0291 & 0.2767 \\ \hline
$1.0000$ & 1.4543 & 0.1410 & 0.0295 & 0.3048 \\ \hline
$1.0500$ & 1.5052 & 0.1348 & 0.0299 & 0.3343 \\ \hline
$1.1000$ & 1.5566 & 0.1291 & 0.0303 & 0.3651 \\ \hline
\end{tabular}
\\ \hline
\end{tabular}
\\[0pt]
Table $5$ (left): Critical values for $q(\varepsilon)=1$, $%
\beta(\varepsilon) =0.1$, $\alpha(\varepsilon) =0.1$ and $d=5$.\\[0pt]
Table $6$ (right): Critical values for $q(\varepsilon)=1$, $%
\beta(\varepsilon) =0.1$, $\alpha(\varepsilon) =0.1$ and $d=6$.
\end{center}

\section{ Closing Remarks}

We have studied black hole solutions of GB--BI gravity's rainbow in
arbitrary dimensions ($d \geq 5$). All GB, BI and gravity's rainbow make
sense in high energy regime and are motivated by the string theory
corrections.

At first, we have given a brief discussion regarding to
geometrical properties and found that obtained solutions are
asymptotically adS with an effective cosmological constant. In
addition, we found that depending the values of parameters, the
black hole singularity may be timelike, spacelike or null. In
other words, such singularity may behave like
Reissner--Nordstr\"{o}m solutions, Schwarzschild black holes or
completely different from them.

Then, we focused on thermodynamical behavior of the solutions. We
calculated conserved and thermodynamic quantities and found that
they may be affected by the existence of rainbow functions and/or
GB parameter. We have shown although BI parameter changes the type
of singularity, it does not change conserved quantities, directly.
Such behavior is expected, since its effect disappears for large
distances ($r \rightarrow \infty$). We have also seen that
although rainbow functions modify conserved and thermodynamic
quantities, the first law of thermodynamics is valid in the energy
dependent spacetime. Then, we calculated the heat capacity and
discussed thermal stability in the canonical ensemble. We found
that depending on the values of free parameters, large black holes
are thermally stable.

Next, we generalized horizon-flat solutions to the case of
rotating and presented exact solutions. Since we used a local
transformation between static and rotating cases, we can confirm
that rotating and static solutions are distinct. We calculated all
related quantities and found that rotation parameters affected all
conserved and thermodynamic quantities. We have also examined the
first law of thermodynamics and proved that its generalization (in
the presence of angular momentum) is already valid.

At last, we have used the extended phase space thermodynamics and
regarded the cosmological constant as a thermodynamical pressure.
We calculated critical quantities and found that depending the
values of free parameters, black hole solutions enjoy a phase
transition.

We have also studied the modified first law of thermodynamics in
the extended phase space with an energy dependent spacetime. We
found that in order to have a consistent Smarr relation and first
law of thermodynamics
in differential form, one has to consider both $\alpha(\varepsilon)$ and $%
\beta(\varepsilon)$ as thermodynamical variables.

As final comment, we should note that in this paper, we have
regarded energy dependent constants without their functional
forms. It is interesting to take into account explicit functional
form of energy dependent constants and investigate their effects.

\begin{acknowledgements}
We gratefully thank the anonymous referee for enlightening
comments and suggestions which helped in proving the quality of
the paper. We also thank S. Panahiyan for reading the manuscript
and useful comments. We also thank Shiraz University Research
Council. This work has been supported financially by the Research
Institute for Astronomy and Astrophysics of Maragha, Iran.
\end{acknowledgements}

\begin{center}
\textbf{Appendix}
\end{center}

In this paper, we tried to present all analytical relations with
general closed forms of rainbow functions. This general closed
forms help us
to follow the trace of temporal ($f(\varepsilon )$) and spatial ($%
g(\varepsilon )$) rainbow functions, separately. However, in order
to plot diagrams and other numerical analysis, we have to choose a
class of rainbow functions. Between all possible choices of
rainbow functions, there are three known classes with various
phenomenological motivations. The first model comes from the hard
spectra of gamma ray burst motivation with the following explicit
forms \cite{RainbowEXP}
\begin{equation}
f(\varepsilon )=\frac{\exp \left( \beta \varepsilon \right) -1}{\beta
\varepsilon },\ \ \ \ \ \ \ \ \ \ g(\varepsilon )=1,  \label{MDR1}
\end{equation}

Taking the constancy of the velocity of light into account, one
can find following relations for the rainbow functions as the
second model \cite{MagueijoSPRL}
\begin{equation}
f(\varepsilon )=g(\varepsilon )=\frac{1}{1-\lambda \varepsilon }.
\label{MDR2}
\end{equation}

Third model is motivated from loop quantum gravity and
non-commutative geometry in which rainbow functions are given as
\cite{Noncom2A,Noncom2B}
\begin{equation}
f(\varepsilon )=1,\ \ \ \ \ \ \ \ \ \ g(\varepsilon )=\sqrt{1-\eta
\varepsilon ^{n}},  \label{MDR3}
\end{equation}%
where $\beta $, $\lambda $, $\eta $ and $n$ are arbitrary
constants which can be fixed according to experimental evidences
near the Planck energy. For numerical calculations, we are
interested in the second model (\ref{MDR2}). In what follows, we
choose suitable values of $\varepsilon $ and $\lambda $ for
obtaining related rainbow functions that used in tables $5$ and
$6$.

\begin{center}
\begin{tabular}{ccc}
\hline\hline $\lambda $ & $\hspace{1cm}\varepsilon
\hspace{1cm}$ & $f(\varepsilon )=g(\varepsilon )$ \\
\hline\hline $-0.5555$ & $0.2000$ & $0.9000$ \\ \hline $-0.2631$ &
$0.2000$ & $0.9500$ \\ \hline $0.0000$ & $0.2000$ & $1.0000$ \\
\hline $0.2381$ & $0.2000$ & $1.0500$ \\ \hline $0.4546$ &
$0.2000$ & $1.1000$ \\ \hline
\end{tabular}%
\end{center}

\end{document}